\documentclass[12pt,a4paper,dvips]{article}
\usepackage{a4p}
\usepackage{cite,mcite}
\usepackage{graphicx}
\begin{document}
%
\vfill
\hfill March 4, 1998
\vfill
\begin{center}
{\Large \bf  
Charge Determination of High Energy Electrons and \\ \vskip 2mm
Nuclei by Synchrotron Radiation with AMS}
\vfill
{\large H. Hofer and M. Pohl\\ \vskip 2mm ETH Z\"urich}
\vfill
{\large \bf Abstract} 
\end{center}
We investigate the possibilities to identify the charge of TeV electrons
and PeV nuclei using their synchrotron radiation in the earth's magnetic field.
Characteristics of synchrotron radiation photons are evaluated and methods of
detection are discussed.
\vfill
\begin{center}
To be published in {\em Nucl. Inst. Meth.}
\end{center}
\vfill
\newpage
%
%
\section*{Introduction}
Cosmic electrons in the TeV energy range are generally assumed to come from
point sources in our galaxy. At high energies between 0.1
and 1 TeV, only a few nearby sources saturate the electron
spectrum~\cite{torii}. Above 1 TeV, there are no measurements.
Several hounded electrons per year are expected above
this energy and inside the acceptance of the Alpha Magnetic Spectrometer, 
AMS~\cite{AMS}.

Most primary positrons would, on the other hand,
annihilate already inside the source, such that only positrons
produced in secondary interactions of charged particles or through primary photons
will arrive at a satellite detector. They would have to be produced
rather close-by through synchrotron light or bremsstrahlung from
ultra-high energy cosmic rays. 
Pioneering experiments using balloons may indicate an excess of positrons
over what is expected by transport models at energies of order 
10 GeV~\cite{barwick,barbiellini,golden}.
Inside the AMS acceptance, one expects several 10 positrons per year with
energies above 1 TeV.

\begin{figure}[htbp]
  \begin{center}
   \includegraphics[width=0.6\textwidth]{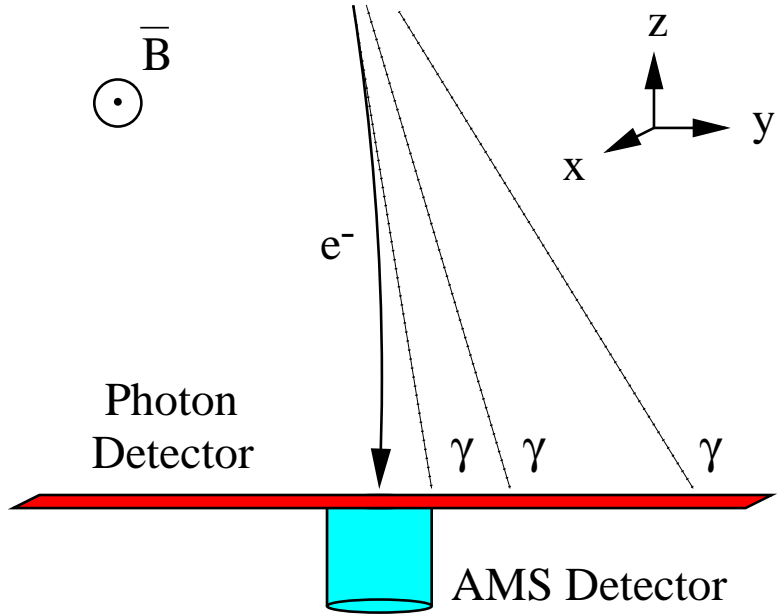}
  \end{center}
  \caption{Sketch of the measurement principle of a synchrotron light detector
  on top of the AMS experiment. For positively charged particles, the synchrotron
  light photons would be detected to the left of the particles trajectory, for
  negatively charged particles to the right. The photon spectrometer will
  be unfolded after installation of AMS on the International Space Station.
  }
  \label{fig:principle}
\end{figure}

For electrons and positrons above about 0.5 TeV impinging on the AMS
detector, a
charge determination and momentum measurement by curvature in the spectrometer
will no longer be possible. Their energy will, however, be measured in the
electromagnetic calorimeter.
In this paper, we discuss how to obtain an
estimate of a particle's momentum and a reliable determination of its
charge sign by using synchrotron light emitted in the earth's magnetic field.
We propose to install a large surface
spectrometer for low energy photons on top of the AMS experiment.
Comparison of the energy measurements from the calorimeter to the one derived
from the number and energy of synchrotron light photons
will allow electron identification.

In addition, ultra high energy nuclei will also produce detectable
synchrotron light. As for electrons, the position of
the synchrotron photons distinguishes particles from antiparticles.
Several events per year are expected in the AMS acceptance from energies
in the knee region of the primary spectrum.

The principle of this measurement is shown in Fig.~\ref{fig:principle}.
Although rather weak, the earth's magnetic field still produces about 20 
synchrotron light
photons per 100 km path length, independent of the electron's energy. When some
of these photons can be detected in an elongated detector aligned with
the magnetic field, photons will be observed on either side of the trajectory
depending on the sign of the particle's charge. Moreover, since the synchrotron
light energy spectrum depends strongly on the primary particle's
energy, the mean energy of the detected photons gives a rough estimate of the 
primary momentum.

In this paper, we calculate the number of photons expected inside a
detector of given length, from an electron with a given momentum
transverse to the earth's magnetic field. We calculate the energy spectrum
of these photons as a function of the primary momentum. 
Finally, we extend the discussion to ultra high energy hadrons.
%
%
\section*{Synchrotron light photons from primary electrons}
In a plane close to the equator, the earth's magnetic field can be approximated
by a vector pointing into the direction of the earth's magnetic poles, with
magnitude
\begin{equation}
B(r) = B_0 \left( \frac{r_0}{r} \right)^3
\end{equation}
with $B_0 \simeq 0.312\ 10^{-4}$T and $r_0 \simeq 6.378\ 10^{6}$m.
In this field, particles with charge $Ze$ follow a helical path with radius of
curvature, $\rho$ given by
\begin{equation}
\rho = \frac{p_{\perp}}{0.3 B Z}
\end{equation}
where $p_{\perp}$ is the component of their momentum perpendicular to $\vec{B}$.
The mean number of synchrotron light photons, $n$,
emitted by a particle with relativistic factor $\gamma = E/m$ on
a path of length $\Delta x$ is given by~\cite{Jackson,Hofmann,Maier}
\begin{equation}\label{equ:Nga}
n = \frac{5 Z^2 \alpha}{2 \sqrt{3}} \frac{\gamma}{\rho} \Delta x
\end{equation}
with the fine structure constant $\alpha \simeq 1/137$.
Since both the relativistic factor and the radius of curvature are
proportional to the particle's energy, this number per unit
length is independent of energy for a given particle species.

\begin{figure}[htbp]
  \begin{center}
  \begin{tabular}{cc}
    \includegraphics[width=0.5\textwidth]{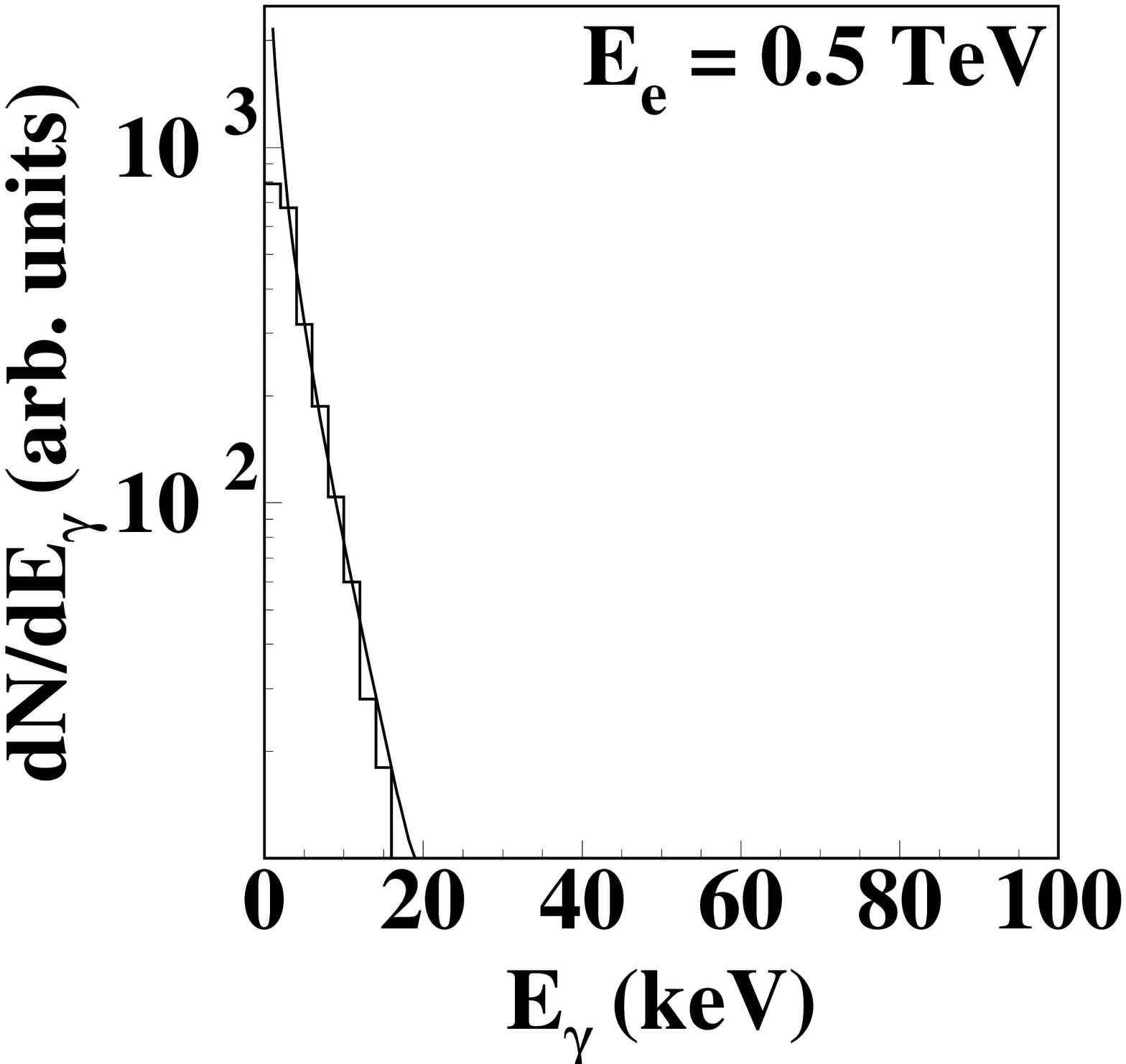} &
    \includegraphics[width=0.5\textwidth]{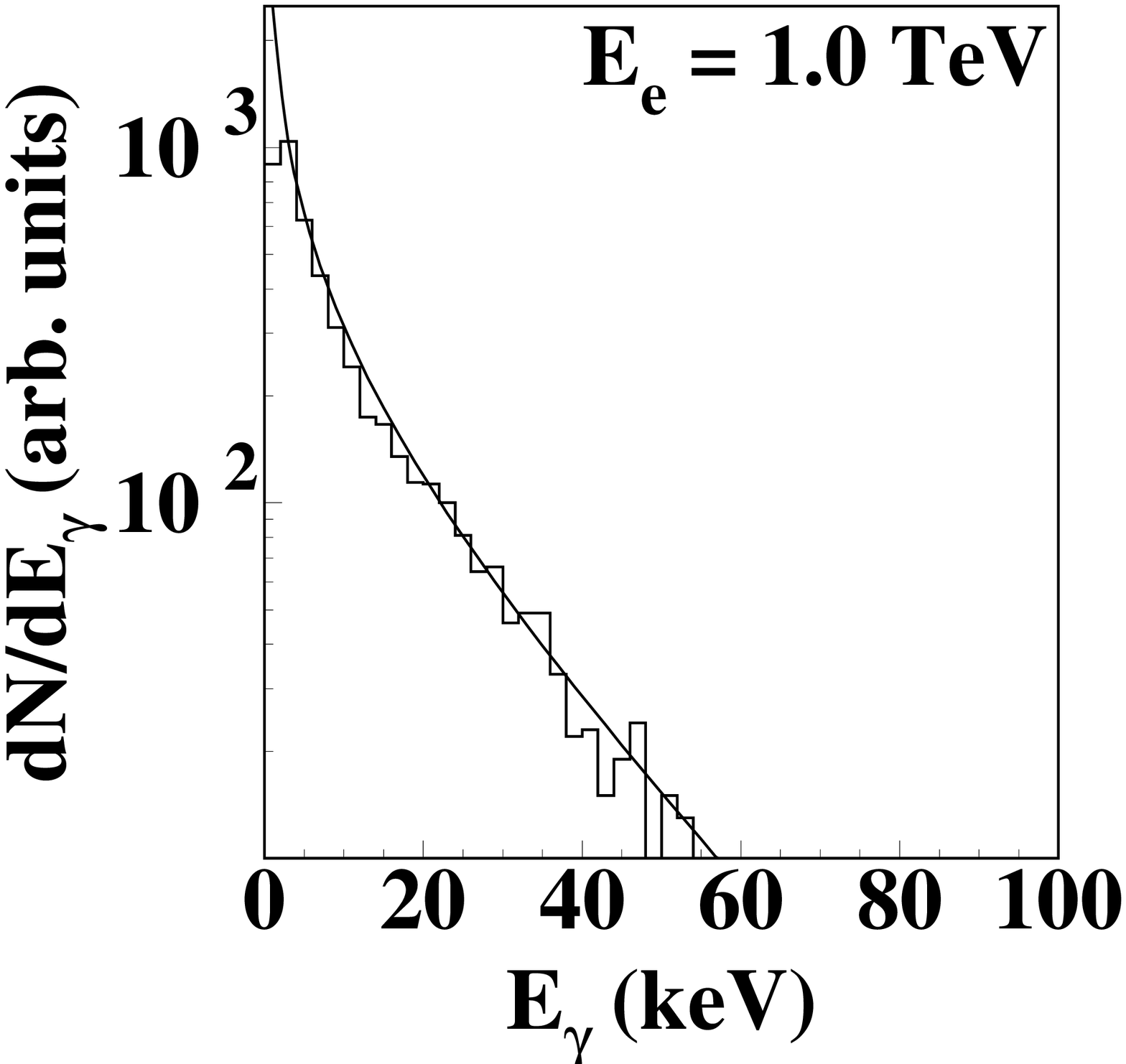} \\
    \includegraphics[width=0.5\textwidth]{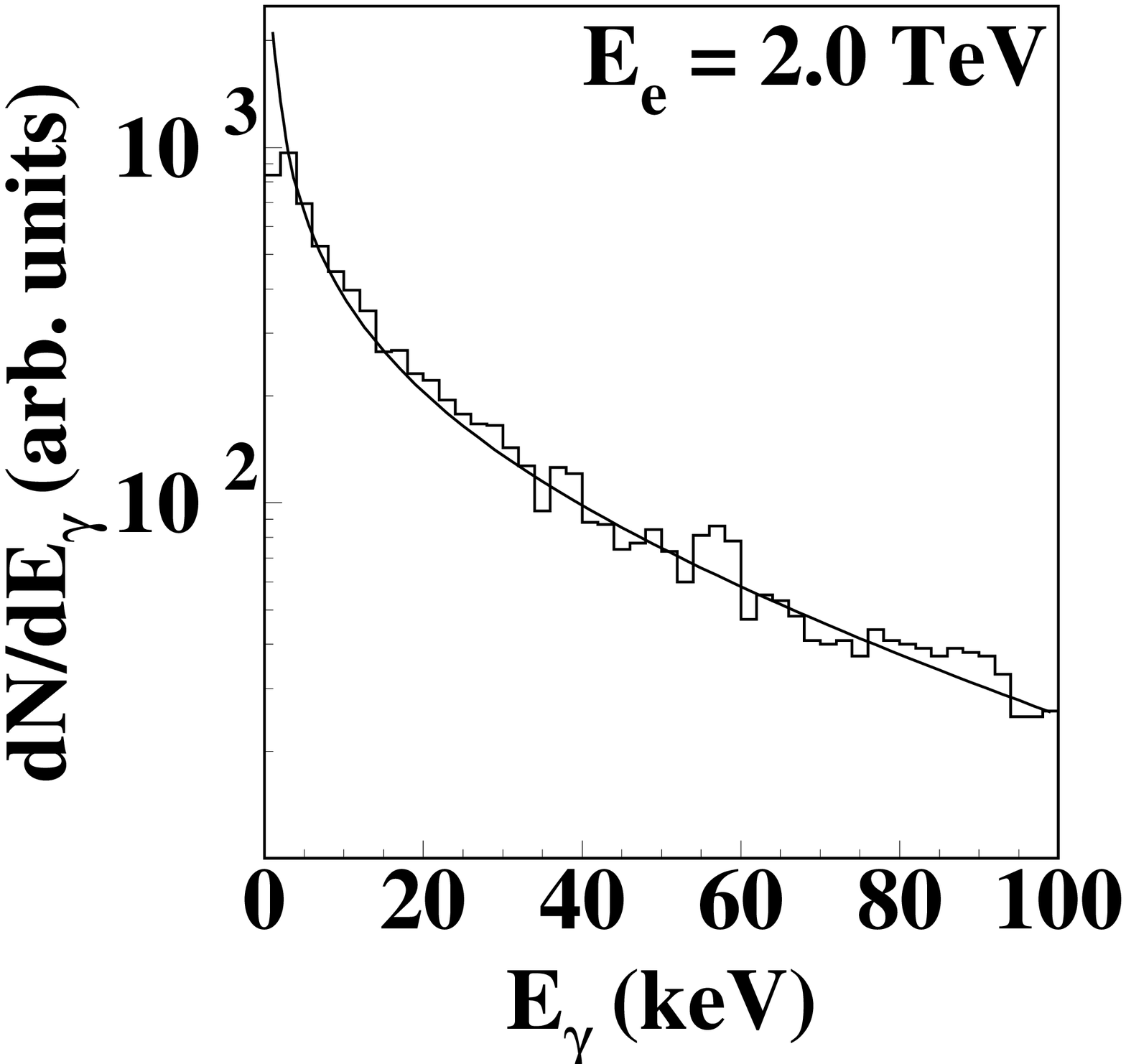} &
    \includegraphics[width=0.5\textwidth]{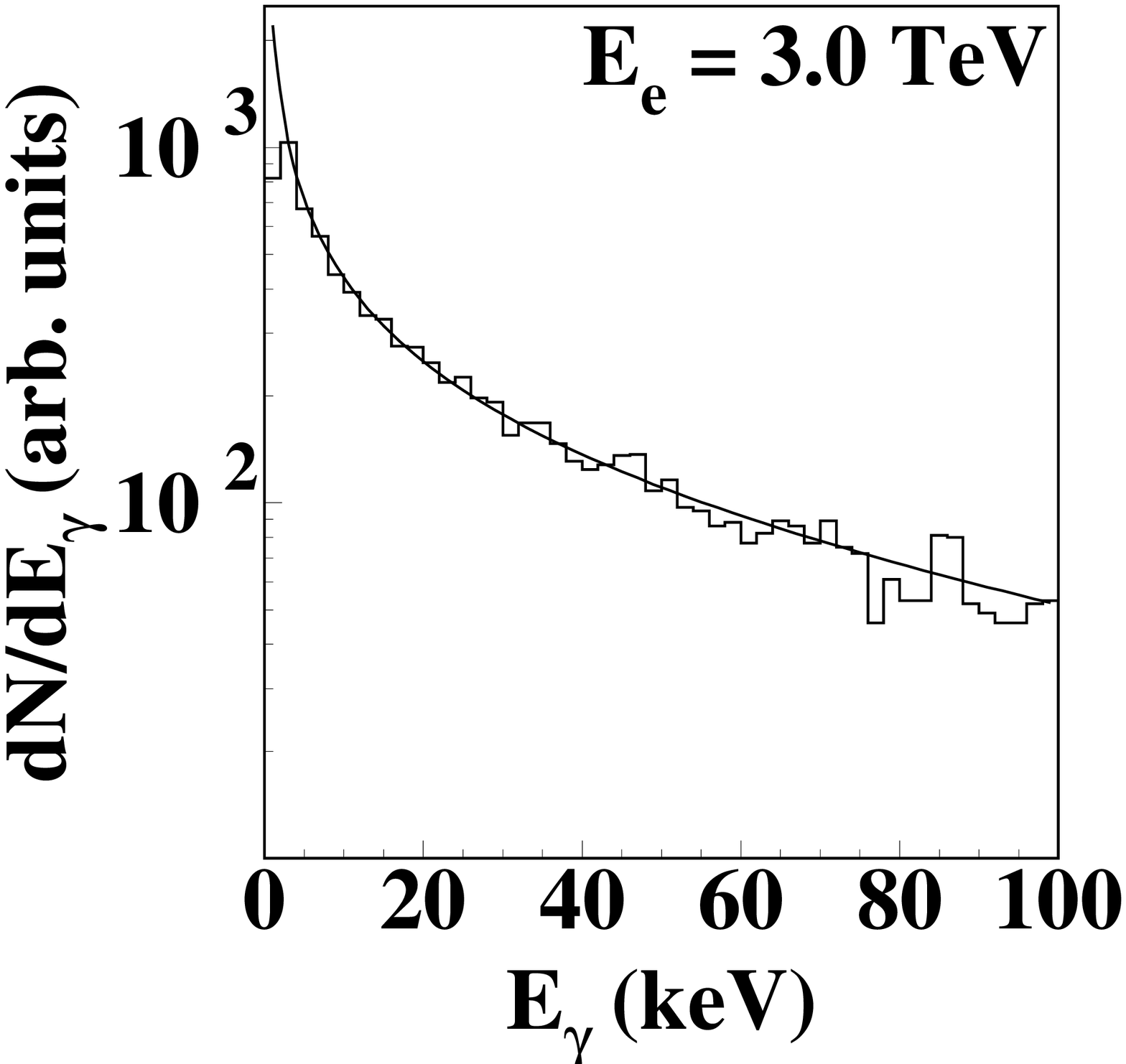}
  \end{tabular}
  \end{center}
  \caption{Spectra of synchrotron light photons for different energies of primary
  electrons. The solid curve is the result of the semi-analytical calculation,
  the histogram is the simulation result with a low energy cut-off of 5 keV.}
  \label{fig:dn_de}
\end{figure}

However, the number of synchrotron photons impinging on a detector
of finite size is not constant with energy.
If one considers a detector of size $\delta y$ normal to
both the particle's momentum vector $\vec{p} = (p, 0, 0)$ and the earth's
magnetic field $\vec{B} = (0, 0, B)$, then photons arrive within $\delta y$ from
path lengths roughly equal to $\Delta x \leq L = \sqrt{\rho^2 - (\rho - \delta
y)^2}$. The effective path length $L$ then multiplies the rate per unit length from
Equ.~\ref{equ:Nga}. A more accurate estimation can be obtained when the
variation of the magnetic field on the path is taken into account.

The energy spectrum of synchrotron light photons also depends critically
on the primary energy. The photon energy spectrum can be expressed
as~\cite{Jackson,Hofmann,Maier}
\begin{equation}\label{equ:spectrum}
\frac{dn}{d\epsilon} = n_0 \int^\infty_{\epsilon/\epsilon_c} K_{5/3}(\xi) d\xi
\end{equation}
with a normalisation constant $n_0$, the modified Bessel function
$K_{5/3}$, the photon energy $\epsilon$ and the characteristic energy $\epsilon_c$
\begin{equation}
\epsilon_c = \frac{3}{2} \hbar c \frac{\gamma^3}{\rho}
\end{equation}
The characteristic energy grows as the square of the primary energy.
The fraction of photons, $f$, above a given energy, $\epsilon_0$,
can be calculated numerically by integrating Equ.~\ref{equ:spectrum} and
normalising the result.
Fig.~\ref{fig:dn_de} shows  photon spectra at electron energies between 0.5 TeV
and 3 TeV. The result of the semi-analytical calculation is compared to
the simulation of synchrotron light using the GEANT Monte Carlo program~\cite{Geant}.
Above a photon energy of 5 keV, a cut-off specified in the simulation runs, 
the spectra agree well.

\begin{figure}[htbp]
  \begin{center}
  \begin{tabular}{cc}
    \includegraphics[width=0.5\textwidth]{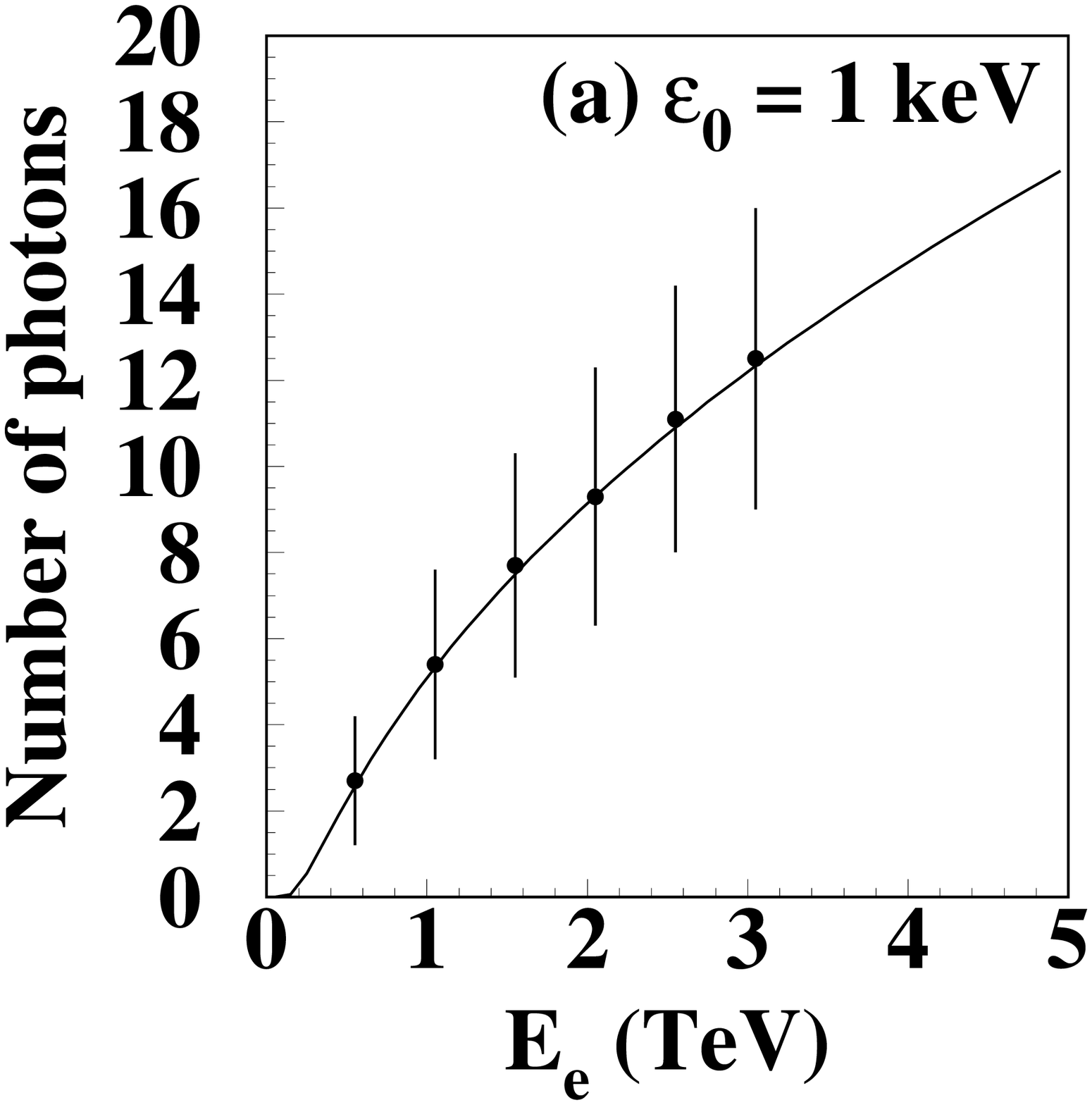} &
    \includegraphics[width=0.5\textwidth]{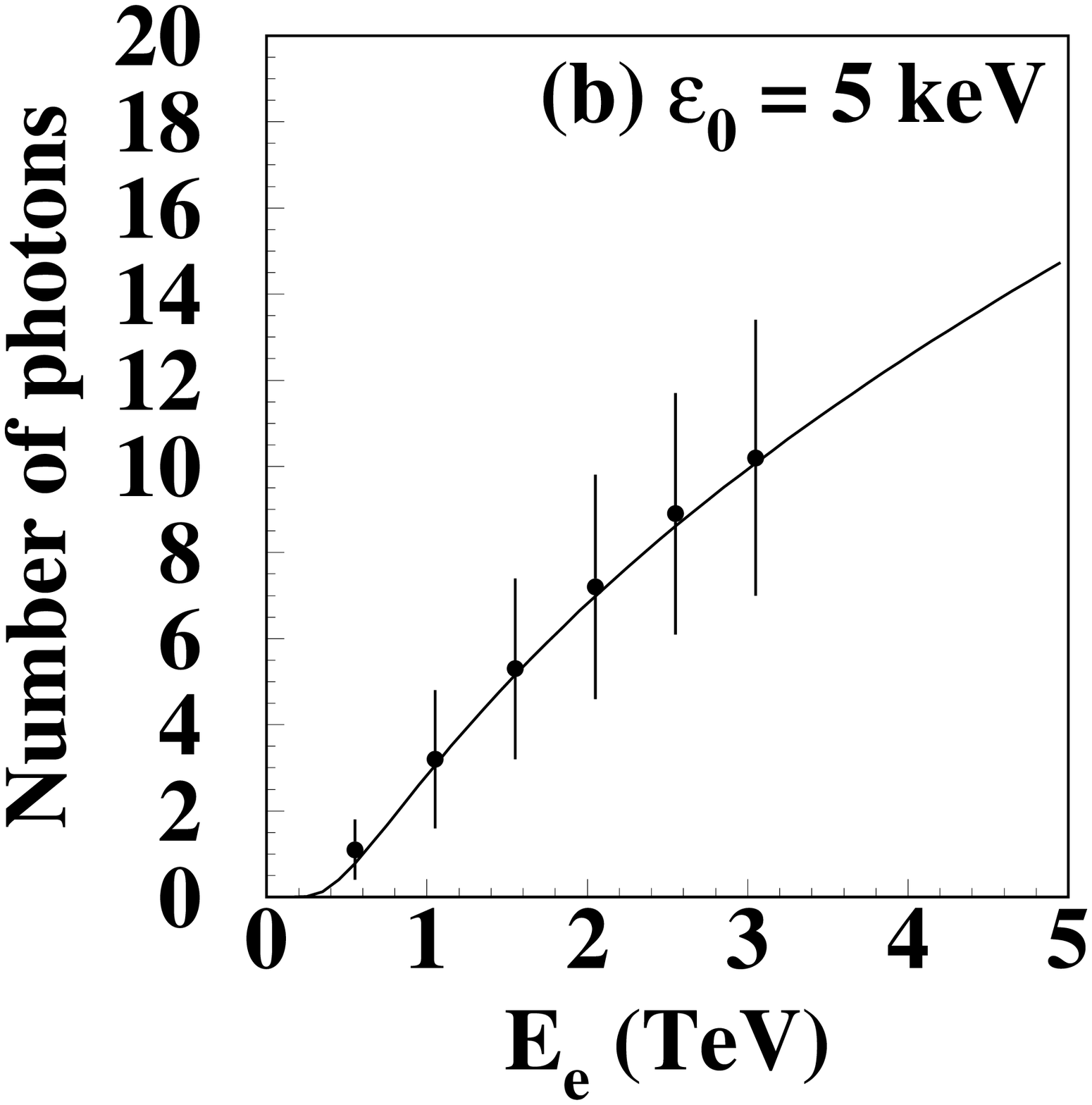}
  \end{tabular}
  \end{center}
  \caption{Mean number of photons impinging on a detector of with $\delta y = \pm 10$ m
  and above a photon energy of (a)$\epsilon_0 = 1$ keV and
  (b)$\epsilon_0 = 5$ keV, as a function of the
  primary electron energy $E_e$. The curve is the result of the semi-analytical
  calculation, the dots indicate the result of a GEANT simulation. The error
  bars denote the width of the number distribution at each energy.}
  \label{fig:Ngam}
\end{figure}

The mean number of photons, $N_\gamma$,
inside a detector of width $\delta y$
and above the minimum detectable energy $\epsilon_0$, is given by
$N_\gamma = (n/\Delta x) L(\delta y) f(E_e,\epsilon_0)$.
The result of the semi-analytical
calculation as well as the result obtained from a simulation of
electron trajectories using the GEANT package~\cite{Geant} are shown in
Fig.~\ref{fig:Ngam}, as a function of the energy of a primary electron.
It is calculated for a very wide detector, $\delta y = \pm 10$ m and a minimum
photon energy $\epsilon_0 = 5$ keV. 
The results of semi-analytical calculation and simulation agree well. 
In a wide detector, between 3 and 10 photons above 5 keV would thus on average be
detected from primary electron energies between 0.5 TeV and 3 TeV.
The number of photons observed in a single event will then be sampled from a
Poisson distribution with expectation $N_\gamma$. The variance of this distribution
is indicated in Fig.~\ref{fig:Ngam} by the vertical error bars.

\begin{figure}[htbp]
  \begin{center}
    \vspace*{-15mm}
    \includegraphics[width=0.70\textwidth]{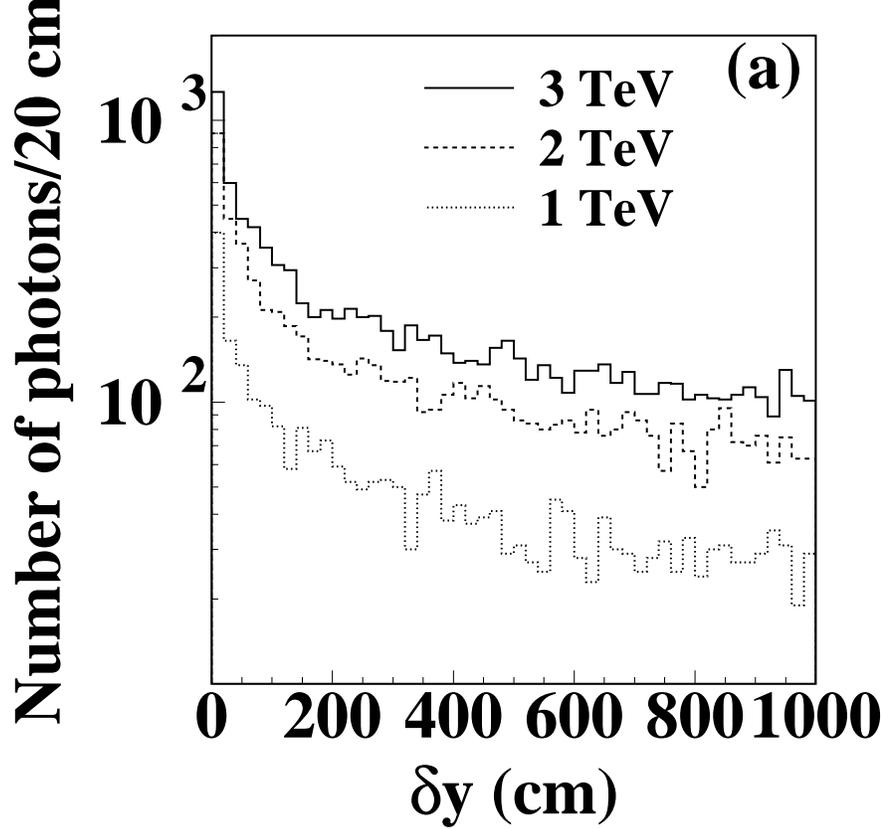}\\
    \vspace*{-10mm}
    \includegraphics[width=0.70\textwidth]{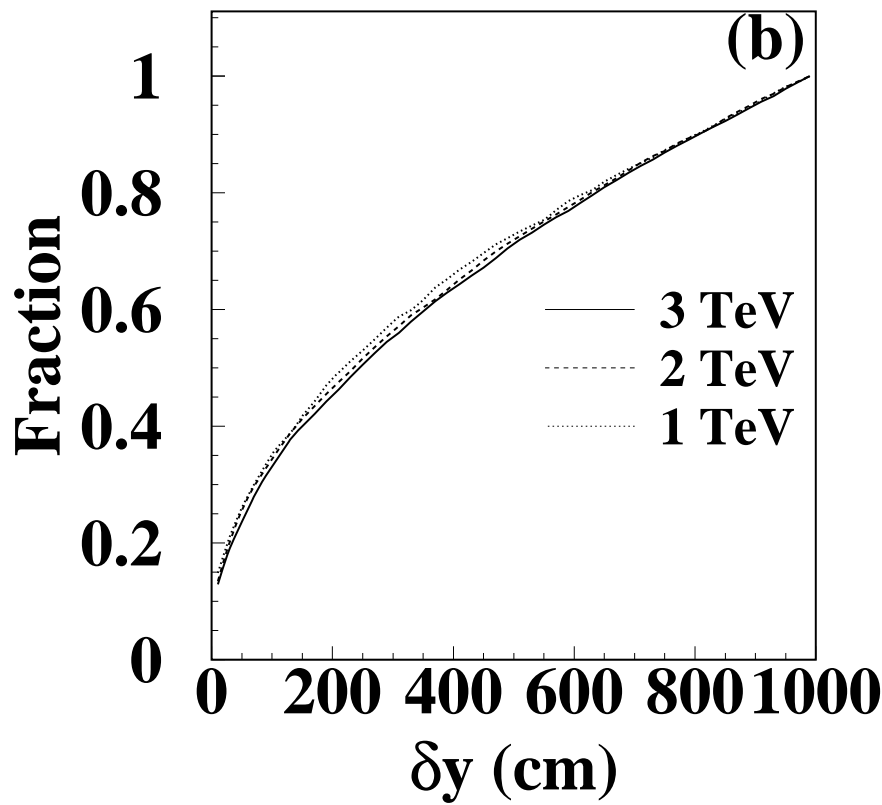}\\
    \vspace*{-5mm}
  \end{center}
  \caption{a) Distribution of the distance $\delta y$
  between the electron and photons impact on the detector, for 1000
  incident electrons at different energies. The shape of the distribution is
  independent of incident energy and photon energy cut off, while the
  number of photons is not.
  b) Fraction of photons impinging on a detector of width $\delta y$,
  relative to $\delta y = \pm 10$ m. This fraction is independent of energy.}
  \label{fig:fraction}
\end{figure}

The fraction of photons inside a
smaller detector is shown in Fig.~\ref{fig:fraction} as a function of
the detector width $\delta y$. This fraction is almost independent of the
primary energy. It is seen that even with a detector as narrow as 
$\delta y = \pm 2$ m,
half of the photons read off Fig.~\ref{fig:Ngam} would still be visible.
The mean number of photons per event would thus be between 1.5 and 5 for a
0.5 TeV to 3 TeV primary electron in a narrow detector.

\begin{figure}[htbp]
  \begin{center}
    \includegraphics[width=0.6\textwidth]{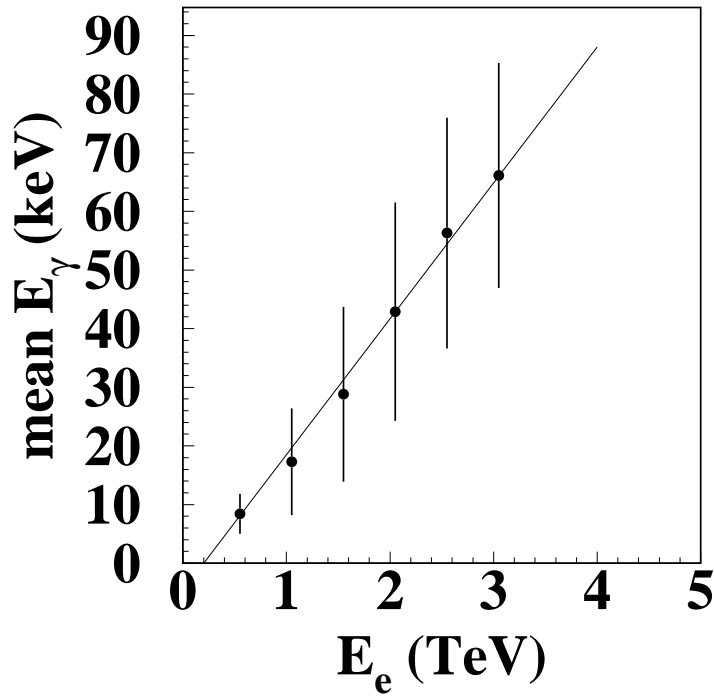}
  \end{center}
  \caption{Mean energy of the photons impinging on the detector above
  a minimum energy of $\epsilon_0 = 5$ keV, as a function of the
  primary electron energy $E_e$. The relation is basically linear.}
  \label{fig:Egam}
\end{figure}

Since both the photon energy spectrum and the number of detected photons
depends crucially on the primary energy,
the number and mean energy of the photons detected in coincidence with the
primary electron or positron can be used to get a rough estimate of the
primary energy. Fig.~\ref{fig:Egam} shows the mean photon energy and the width
of its distribution as a function of the primary energy. The dependence is
almost linear. The resolution will be of the order of 30\% from this measurement
alone. In addition, a measurement with similar resolution will follow from
the relation between the number of photons and the primary energy, as shown in
Fig.~\ref{fig:Ngam}. Combining the two will thus result in an improved
measurement, with a total resolution of order 20\%. For electrons and positrons
the energy derived from the number and mean energy of synchrotron light photons
should be equal to the energy measured in the calorimeter, which has much better
resolution. A reliable separation of electrons from nuclei should thus be
possible.

\section*{Synchrotron radiation from nuclei}

Low energy synchrotron light is also emitted by ultra high energy nuclei.
The number of synchrotron
photons emitted per unit path length is proportional to
$Z^3/M$, where $Z$ is the nucleus' charge and $M$ is its mass, and independent
of energy. The path length contributing to photons impinging on the
detector roughly scales like $\sqrt{E/Z}$, where $E$ is the energy of the nucleus,
such that energies in the PeV range give observable photon rates. 
The photon spectrum is characterised by $\epsilon_c \sim E^2 Z / M^3$ and
thus in the eV range for the lighter nuclei and PeV energies.

\begin{figure}[htbp]
  \begin{center}
    \includegraphics[width=0.90\textwidth]{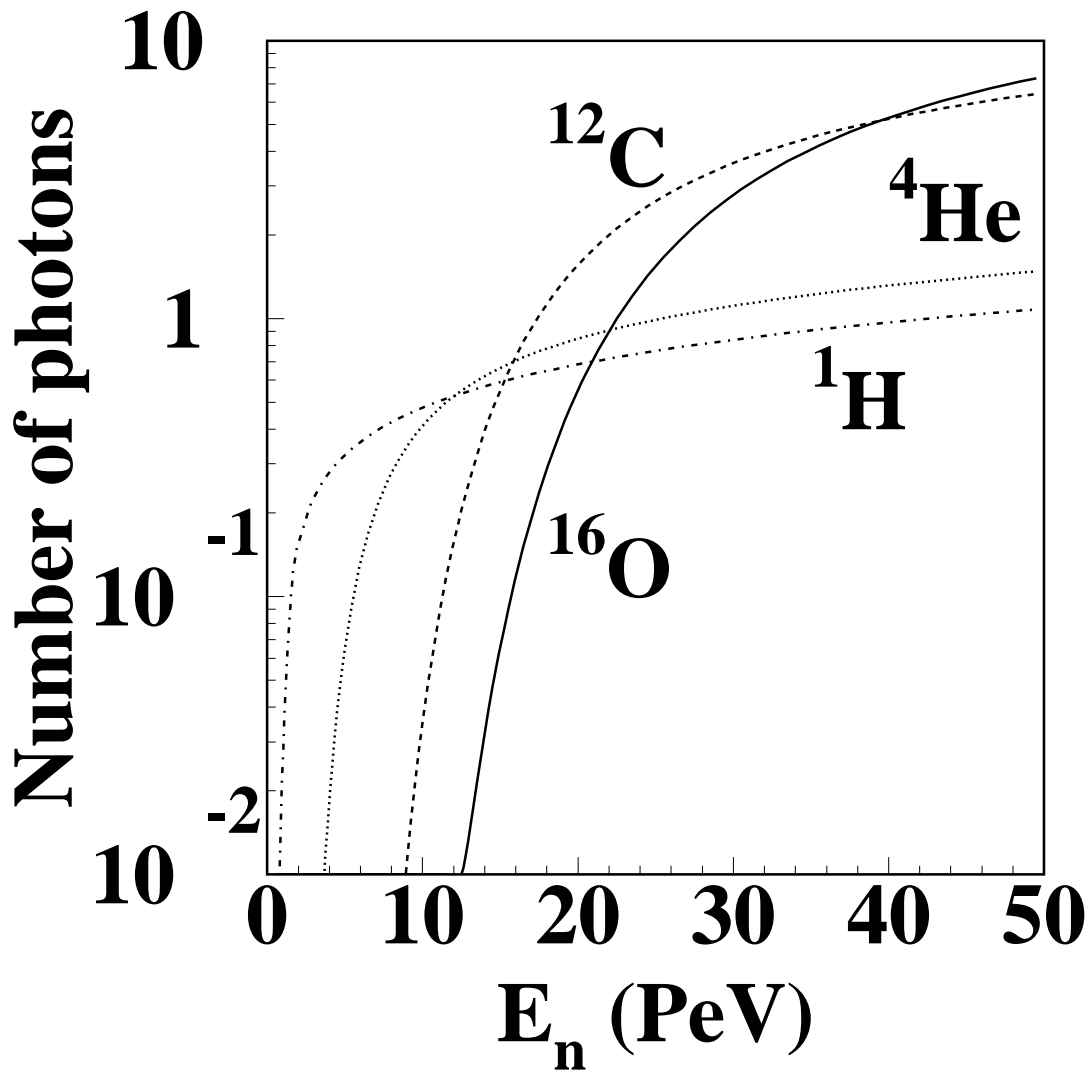}
  \end{center}
  \caption{Mean number of photons impinging on a detector of width $\delta y = \pm 10$ m
  and above a photon energy of $\epsilon_0 = 5$ eV, as a function of the
  primary nucleus energy $E_n$.}
  \label{fig:Ngamn}
\end{figure}

Fig.~\ref{fig:Ngamn}
shows the expected number of photons above 5 eV from protons and light
nuclei. The geometrical suppression is not expected to change with
respect to what is shown in Fig.~\ref{fig:fraction}. Thus, measurable
signals are expected for nuclei from the knee region of the energy spectrum.
This requires a detector sensitive to photons of a few eV energy.

\section*{Conclusions}
Observable synchrotron light is emitted by TeV electrons and PeV nuclei in the
earth's magnetic field. Its position with respect to the primary momentum
and the field direction can be used to measure
the sign of the particle charge and thus distinguish particles from antiparticles.
Moreover, counting the number of synchrotron light photons and measuring their
mean energy, an estimate of the primary electron momentum is obtained.
This estimate can be used to distinguish electrons from nuclei.

Synchrotron light can be detected and its energy measured using 
a photon spectrometer of several square meters surface mounted on top of the
AMS experiment. It must be sensitive to photon energies above a few keV for
detection of an electron signal, above a few eV for detection of a signal from
nuclei. Candidate 
technologies for such devices are solid state as well as gaseous and crystal
photon detectors.

\section*{Acknowledgements}

We wish to express our gratitude to Prof.~A.~Hofmann, Dr.~P.~Le Coultre
and Prof.~A.D.~Erlykin for clarifying discussions.

\end{document}